\documentclass{article}
\usepackage[utf8]{inputenc}
\usepackage{authblk}
\usepackage[dvips]{graphicx}
\graphicspath{{noiseimages/}}
\usepackage{xcolor}
\usepackage[T2A]{fontenc}
\usepackage{tikz}
\usepackage{cite}
\usepackage{multirow}
\usepackage{pgfplots}
\usepackage{mathrsfs}

\pgfplotsset{compat=1.7}
\usetikzlibrary{intersections, pgfplots.fillbetween}
\usetikzlibrary{decorations.pathmorphing}
\usetikzlibrary{arrows.meta}
\usepackage[mode=buildnew]{standalone}
\usepackage[toc,page]{appendix}
\usepackage{amsmath}
\usepackage{physics}
\usepackage{amssymb}
\usepackage{float}
\usepackage{comment}
\usepackage{a4wide}
\usepackage[english]{babel}
\usepackage{hyperref}
\definecolor{urlcolor}{HTML}{990000}
\definecolor{linkcolor}{HTML}{005F5F}
\hypersetup{pdfstartview=FitH,  linkcolor=linkcolor,urlcolor=urlcolor, colorlinks=true,citecolor=blue}
\usepackage[page]{appendix}

\usepackage[english]{babel}
\usepackage{hyperref}
\usepackage[T2A]{fontenc}
\usepackage[utf8]{inputenc}

\author[1,2]{E.~T.~Akhmedov\thanks{\href{mailto:akhmedov@itep.ru}{akhmedov@itep.ru}}}
\author[1]{I. A. Belkovich\footnote{\tt belkovich.ia@phystech.edu }}
\author[1,3]{D. V. Diakonov\footnote{\tt dmitrii.dyakonov@phystech.edu}}
\author[1]{K. A. Kazarnovskii\footnote{\tt kazarnovskiy.ka@phystech.edu}}
\affil[1]{\itshape Institutskii per, 9, Moscow Institute of Physics and Technology, 141700, Dolgoprudny, Russia}
\affil[2]{\itshape Academician Kurchatov Square, 1, NRC ''Kurchatov Institute'', 123182, Moscow, Russia}
\affil[3]{\itshape Bol'shoi Karetnyi per., 19, Institute for Information Transmission Problems, 127994, Moscow, Russia}
\title{\textcolor{black}{On (dis)agreement between different methods of calculation of the imaginary part of the effective action in expanding space-times}}
\begin{document}

\numberwithin{equation}{section}
\date{}
\maketitle
\begin{abstract}
    We consider two approaches to calculate imaginary parts of effective actions in expanding space-times. While the first approach uses Bogolyubov coefficients, the second one uses the functional integral or the Feynman propagator. In eternally expanding space-times these two approaches give different answers for the imaginary parts. The origin of the difference can be traced to the presence if the wave-functionals for the initial and final states in the functional integral. We show this explicitly on the example of the expanding Poincare patch of the de Sitter space-time.
\end{abstract}

\section{Introduction}
\quad  The phenomenon of particle creation by external fields from the vacuum is of great interest and is widely discussed in the literature, see e.g. \cite{Parker:1968mv, Parker:1969au, Zeldovich:1971mw, Zeldovich:1970si, Grib:1980aih, Birrell:1982ix, Cespedes:1989kh, Ford:2021syk}. The process of converting the external field and state of the theory into ordinary matter are important in the large number of fundamental quantum effects such as Hawking radiation \cite{Hawking:1975vcx,Gibbons:1977mu}, the Unruh effect \cite{Unruh:1976db}, Schwinger mechanism \cite{Schwinger:1951nm} and cosmological particle creation \cite{Gibbons:1977mu} and etc. 

In cosmology, the process of particle creation may play an important role in the mechanism of reheating after inflation \cite{Kofman:1994rk,Kofman:1997yn} which is believed to be responsible for the creation of almost all the particles in our Universe (see also \cite{Akhmedov:2017dih}). Apart from this the large amount of entropy of our universe can be explained by the cosmological pair creation from the initial state \cite{Gasperini:1992xv}.

In general particle creation occurs on time dependent backgrounds even for the fields without self-interactions. In such a case the free Hamiltonian is explicitly time-dependent and not diagonal. In such a situation the initial vacuum state is no longer vacuum in the future but corresponds to an excited state. Which is, in fact, interpreted as the process of particle creation. 


In this paper we consider the process of particle creation only in the Gaussian approximation (see e.g.\cite{Akhmedov:2011pj,Akhmedov:2013vka} for the situation in loops for the backgrounds under consideration) to clarify the following discrepancy.
There are a two commonly used methods (we will describe them in details below) to compute the probability of particles creation. However, it turns out that in some cases in curved space-times these two methods give different answers. For example, in de Sitter space-time the method that uses coefficients of the Bogolyubov transformation (between in- and out-states) to calculate the probability of particle creation gives the following answer \cite{Mottola:1984ar,Anderson:2013ila}:  
\begin{equation}
W =\frac{V_3}{2} 
 \int \frac{d^3 p}{(2\pi)^3} \ln |\alpha_p|= -V_4 H^4 \frac{\ln{\left(1-e^{-2\pi\sqrt{(m/H)^2-9/4}}\right)}}{(2\pi)^4},
\end{equation}
where $m$ is the mass of the scalar field, $V_3=\int d^3x$ is the spatial volume, $V_4$ is the space-time volume and $\alpha_p$ is one of the coefficients of the Bogolyubov transformation:
\begin{equation}
    b_p=\alpha_p a_p+\beta_p a^+_p,
\end{equation}
between the ladder operators, $a,a^+$ and $b,b^+$, for ``in'' and ``out'' modes.

At the same time the probability of the particle creation can be expressed in terms of the imaginary part of the Feynman in-out propagator at coincident points\cite{Candelas:1975du,Polyakov:2007mm,Akhmedov:2009ta,Akhmedov:2019esv}  :
\begin{align}
\label{im ds G}
W = - \frac{1}{2} \int d^{4} x \sqrt{|g|} \int\limits_{\infty}^{m^2} d \tilde{m}^2 \Im G(x,x) \approx V_4\frac{H^4}{16 \pi^{2}} \left(\frac{m}{H}\right)^{3} e^{- 2 \pi m/H},\quad \text{when} \quad m\gg H.
\end{align}
One can see that these two methods give different answers. It is important to note that these methods are non-perturbative in the intensity of the external field. Let us point out here that there is also another method for computing the probability of particle creation \cite{Frieman:1985fr,Campos:1993ug,Dobado:1998mr,Akhmedov:2024axn}, which, depending on conditions, can give perturbative or non-perturbative answers. For a comparison of the results \eqref{im ds G} of particle creation with perturbation theory in de Sitter spacetime, see \cite{Zhou:2024oxe}.

In this paper, we will show that if the space-time expansion takes a finite time, then the two methods described above give the same answer, but in the case of the infinite expansion, they disagree. We will also discuss how this fact is related to the definition of the functional integral. 

The paper is organized as follows. In Section 2 we give an overview of the method to calculate the in-out amplitude in terms of Bogolyubov coefficients that relate the ''in'' and ''out'' positive energy modes. In Section 3 we describe the method of calculating the in-out amplitude in terms of the functional integral and show that the effective action can be expressed in terms of the Feynman in-out propagator at coincident points. In Section 4  we show that both method give the same answer if a space-time is expanding for a finite period of time. Then we show that in the case of infinite space-time expansion these two methods give different answers. In Section 5 we give several examples when these two methods give different answers. In Section 6 we argue why these two methods give different answers on the example of the expanding Poincare patch of the de Sitter space-time.

\section{In-out amplitude from the Bogolyubov coefficients}

In this article we discuss phenomenon of the particle creation on the example of massive scalar particles in four dimensional curved space-times. We consider the theory of minimally coupled real scalar field theory defined by the action:
\begin{align}\label{action}
    \mathcal{S}=\frac{1}{2}\int d^4 x \sqrt{g} \Big(\partial_\mu \phi(x) \partial^\mu \phi(x)-m^2\phi^2(x)+\xi R \phi^2(x) \Big).
    \end{align}
For simplicity, we restrict our consideration to spatially homogeneous and isotropic cosmological space-times:  
\begin{equation}
\label{metric}
    d s^2 = \Omega^2(t) (d t^2 - d  \vec{x}^2),
\end{equation}
with the Ricci scalar: 
\begin{gather}
 \quad R= -\frac{6\partial_t^2\Omega(t)}{\Omega^3(t)}.
\end{gather}
The field operator is the solution of the Klein-Gordon equation: 
\begin{align}
\label{KG}
    \left(\Box+m^2-\xi R\right) \phi(x)=0
\end{align}
and can be expanded in terms of ''in'' or ''out'' modes:
\begin{align}
\label{fieldoperator}
    \hat{\phi}(x)=\int \frac{d^3 p}{(2\pi)^{3/2}}\frac{1}{\Omega(t)} \Bigg(  \varphi^{in / out}_p(t) e^{-i p x} \hat{a}^{in / out}_p+h.c. \Bigg),
\end{align}
where $\varphi^{in}_p(t)$ and $\varphi^{out}_p(t)$ are the solution of the following equation:
\begin{align}
\label{mode}
    \partial_t^2\varphi^{in / out}_p(t) +\left[p^2+\Omega^2(t)\left(m^2+\left(\xi-\frac{1}{6}\right)\frac{6\partial_t^2\Omega(t)}{\Omega^3(t)}\right)\right]\varphi^{in / out}_p(t) = 0,
\end{align}
 and the creation and annihilation operator satisfy the standard commutation relations: 
\begin{align}
  \quad  \left[\hat{a}^{in / out}_p,\left(\hat{a}^{in / out}_k\right)^\dagger\right]=\delta^3(p-k).
\end{align}
Expansion \eqref{fieldoperator} defines the ''in'' and ''out'' Fock space ground states:
\begin{align}
   \hat{a}^{in}_p |in\rangle =0 \quad \text{and} \quad \hat{a}^{out}_p |out\rangle =0.
\end{align}
These modes and ladder operators are related by Bogolyubov transformations: 
\begin{align} \label{bogolyubov}
\varphi^{out}_p(t)=\alpha^*_p \varphi^{in}_p(t)-\beta_p \varphi^{in*}_p(t) \quad \text{and} \quad  \hat{a}^{out}_p=\alpha_p \hat{a}^{in}_p+ \beta_{p}^*\hat{a}^{in \dagger}_{-p}.
\end{align}
In general in eternally expanding space-times the ''in'' and ''out'' modes are defined without imposing the positive energy conditions. But in space-times with  $\Omega(t)$ which tends to constants at past and future infinities one can define the ''in'' and ''out'' modes as positive energy solutions at past and future infinities as follows: 
\begin{align}\label{inassym}
      \varphi^{in}_p(t\to -\infty)\approx \frac{1}{\sqrt{2 \omega_{p-}}} e^{-i \omega_{p-} t}  \quad\text{and}   \quad  \varphi^{in}_p(t\to +\infty)\approx \frac{1}{\sqrt{2 \omega_{p+}}}\Bigg( \frac{1}{T_p} e^{-i \omega_{p+} t}+ \frac{R_{Lp}}{T_p} e^{i \omega_{p+} t} \Bigg)
\end{align}
and
 \begin{align}
 \label{assymptoticb}
   \varphi^{out}_p( t\to +\infty)\approx \frac{1}{\sqrt{2 \omega_{p+}}} e^{-i \omega_{p+} t} \quad \text{and}  \quad  \varphi^{out}_p( t\to -\infty)\approx \frac{1}{\sqrt{2 \omega_{p-}}}\Bigg( \frac{1}{T_p^*} e^{-i \omega_{p-} t}+ \frac{R^*_{Rp}}{T_p^*} e^{i \omega_{p-} t} \Bigg).
\end{align}
The reflection ($R$) and transition ($T$) amplitudes obey the relations:
\begin{align}
    R_{Rp} T_p^*= - R_{Lp} T_p \quad \text{and} \quad |R_{R,Lp}|^2+|T_p^2|=1,
 \end{align}
which follow, e.g., when one establishes simultaneously the canonical commutation relations between the ladder operators and between the field operator with its conjugate momentum.
These $R$ and $T$ amplitudes can be expressed in terms of the Bogolyubov coefficients between in- and out-modes:
\begin{align}
    T_p=\frac{1}{\alpha_p}, \quad R_{Rp}=-\frac{\beta_p^* }{\alpha_p} \quad \text{and} \quad  R_{Lp} = \frac{\beta_p^* }{\alpha^*_p} .
    \end{align}
    
To calculate the in-out transition amplitude, $\langle out|in\rangle$, we can formally express the state $|in\rangle$ in terms of  the out-Fock space \cite{Parker:1969au,Ford:2021syk}:
\begin{gather}
    |in\rangle =\left(\prod_{p,\  p_z \geq 0} |\alpha_p |\right)^{-1} \exp\left[-\sum_{p,\  p_z \geq 0} \left(\frac{\beta^*_p}{ \alpha_p}  \hat{a}^{out \dagger}_p \hat{a}^{out \dagger}_{-p} \right)\right]  |out\rangle
    \nonumber
    =\\= \left(\prod_{p,\  p_z \geq 0} |\alpha_p |\right)^{-1} \sum_{ \{n_p\}} \prod_{p,\  p_z \geq 0}\left(\frac{\beta^*_p}{\alpha_p^*}\right)^{n_p} |\{n_p\}\rangle.
\end{gather}
Here we denote $ \{n_p\} $ as a set of occupation numbers i.e., for each mode $p$ there is $n_p$ which gives the number of particles with momentum $p$, as well as $n_p$ particles with momentum $-p$ over the out vacuum. To avoid overcounting in the product over the momentum in the last equation we restrict $p_z\geq0$ \cite{Ford:2021syk}. Indeed to prove the last relation one can use the Bogolyubov transformation to check that: 
\begin{gather}
     \hat{a}^{in}_k |in\rangle
     \nonumber=\\=
     \left(\alpha^*_k \hat{a}^{out}_k- \beta_{k}^*\hat{a}^{out \dagger}_{-k}\right)  \left(\prod_{p,\  p_z \geq 0} |\alpha_p |\right)^{-1} \sum_{ \{n_p\}} \prod_{p,\  p_z \geq 0}\left(\frac{\beta^*_p}{\alpha_p^*}\right)^{n_p} |\{n_p\}\rangle =0.
\end{gather}
As a result the amplitude for the ''in'' state to remain the ''out'' state is as follows: 
\begin{align}
\label{outin}
   \langle out|in\rangle= \left(\prod_{p,\  p_z \geq 0} |\alpha_p |\right)^{-1}.
\end{align}
If the probability for the in-state to remain the out-state is equal to $1$, then the background field under consideration does not create particles (at least in the gaussian approximation). Hence the ''in'' and ''out'' states can be identified up to a phase. Such a situation occurs in the case when the potential in the Schroedinger type equation \eqref{mode} is reflectionless (when the reflection coefficient is vanishing). For example, this happens in the odd dimensional de Sitter space where the potential is a solution of the KdV equation which gives reflectionless potential (see e.g. \cite{Das:2006wg,Polyakov:2007mm,Akhmedov:2019esv}) and one can define some peculiar ``in'' and ``out'' states in global coordinates \cite{Mottola:1984ar}. 

If the probability is less than $1$, it usually interpreted as the sign of the particle creation. The total probability  of the particle creation can be defined as follows:
\begin{gather}
\label{WB}
W_{B}=-\ln | \langle out|in\rangle|=\int d^4 x \sqrt{g}w_B(x)=N \frac{V_3}{2} 
 \int \frac{d^3 p}{(2\pi)^3} \ln |\alpha_p|
\end{gather}
where $w_B(x)$ is the probability rate per unit four volume, $V_3$ is the volume of space and index $B$ means that we express the probability in terms of Bogolyubov coefficients, $N$ is the number of independent scalar fields, (one for a real field or two for a complex charged scalar), additional $\frac{1}{2}$ arises from the restriction $p_z\geq0$, which we introduced above to avoid doublecounting.  

The last integral in \eqref{WB}  is well defined in the case when the background field is switched on and then switched off adiabatically. In such a case the particle creation occurs during a finite period of time. Hence the rate per unit volume $W_B/V_3$ is finite. 

At the same if the external field acts eternally, then the rate per unit volume $W_B/V_3$ diverges. In such a case the divergent integral over momentum in \eqref{WB} can be associated with the infinite duration of the action of the background field and allows one to define the finite rate of particle creation $w(x)$ per unit time and volume. For example, one can consider creation process of charged scalars in constant electric field in the time dependent gauge $A_\mu=(0,0,0,-E t)$. In this case the Bogolyubov coefficient is $|\alpha_p|^2=1+\exp{\left(\frac{-\pi(m^2+p^2_{\perp})}{|e|E}\right)}$ and is independent of $p_z$ \cite{1969JETP...30..660N,Grib:1980aih,Anderson:2017hts}. Therefore the integral over $p_z$ in \eqref{WB} is linearly divergent. Then, using the expression for the physical momentum, $p_{phys}=p_z+e E t$, one can associate the linear divergence in the integral over $p_z$ with the infinite duration of time: $\int dp_z= - e E \int dt $ . As a result the probability of the particle creation is given by: 
\begin{gather}\label{Schwingerresult}
   W_B   = V_{3} T \frac{|e|^2 E^2}{2 (2\pi)^3 } \sum_{n = 1}^{\infty}  \frac{(-1)^{n+1}}{n^2}e^{-\frac{\pi m^2 n}{|e|E}}. 
\end{gather}
The total probability is proportional to the space-time volume $V_4=V_3T$ and the probability is constant in space and time, which should be the case for the eternal and constant background field.

This identification of the infinities in the divergent integrals over momentum and over the time appears to be non-universal. As we will demonstrate in the next section, if we attempt to determine the rate of particle creation over a finite time period (but when the background field is still on), we will also encounter a divergent integral over momentum that cannot be identified with a finite duration of time.

Moreover, as we just pointed out, in the constant electric field background the rate of particle creation is finite and homogeneous, $w(x)=const$. However, for a general cosmological space-times described by \eqref{metric}, we expect that the rate of particle creation can be a function of time $w(t)$. Therefore, the divergent integral over the momentum should be identified not only with the divergent integral over the time, but should also account for the fact that the rate of particle creation may depend on time. Consequently, this method is applicable only in cases when the external field is turned off at both past and future infinities, ensuring that the rate of particle creation remains finite.

\section{Particle creation during finite period of time when the background field is still on} 
In this section, we show explicitly that the correct definition of the rate of particle creation depends on how we define the Hamiltonian operator in the theory. Namely, we calculate the rate of particle creation between two instantaneous Fock space ground states at \(t=t_i\) and \(t=t_f\). These states are defined by the diagonalization of the corresponding free Hamiltonian at the given moments in time. The instantaneous Fock space ground state does not contain any particles at \(t_i\), but at another time, it corresponds to an excited state. We consider three different definitions of the Hamiltonian \cite{Fulling:1979ac} and discuss the consequences. 


If the canonical variables are \(\phi\) and \(\pi = \frac{\partial \mathcal{L}}{\partial \dot{\phi}}\), then the canonical Hamiltonian operator in the theory (\ref{action}) has the standard form: 
\begin{gather}
\label{Ham}
    \hat{H}_\phi=\int d^3x \left(\partial_t \hat{\phi}(x)\hat{\pi}(x) - \mathcal{L}\right) 
    =\\=
    \frac{1}{2} \int d^3x \sqrt{g} \Bigg( \nonumber  g^{00}\partial_t \hat{\phi}(x)\partial_t \hat{\phi}(x) -g^{i j}\triangledown_i\hat{\phi}(x)\triangledown_j\hat{\phi}(x)+\left(m^2-\xi R\right)\hat{\phi}^2(x)\Bigg),
\end{gather}
where the canonical momentum is given by \(\pi=\sqrt{g} g^{00}\partial_t\phi\) and the index \(\phi\) of $ \hat{H}_\phi$ means that we use \(\phi\) as the canonical variable. 

The commutation relations of the field operator and the conjugate momentum with the canonical Hamiltonian have the standard form: 
\begin{align}
    \left[\hat{H}_\phi,\hat{\phi}\right]=-i \dot{\hat{\phi}} \quad \text{and} \quad \left[\hat{H}_\phi,\hat{\pi}\right]=-i \dot{\hat{\pi}},
\end{align}
and lead to the standard equations of motion. Namely, using the last equations, one can show that the field operator satisfies the Klein-Gordon equation \eqref{KG}. This trivial observation will be important for the discussion below. 

The expression for the canonical Hamiltonian via the ladder operators is: 
\begin{align}
     \hat{H}_\phi=\frac{1}{2}\int \frac{d^3 p}{(2\pi)^3} \omega_\phi(p,t) \bigg[ K_\phi(p,t) \hat{a}_p \hat{a}_p^\dagger + \Lambda_\phi(p,t) \hat{a}_p {\hat{a}_{-p}} \bigg]+h.c.,
\end{align}
where:
\begin{align}
   \omega^2_\phi(p,t)=p^2+\Omega^2(t)\left(m^2+\xi\frac{6\ddot{\Omega}(t)}{\Omega^3(t)}\right),
\end{align}
\begin{align}
K_\phi(p,t)=\frac{1}{\omega_\phi(t)}\left[\left|\dot{\varphi}_p(t)-\varphi_p (t)\frac{\dot \Omega}{\Omega}\right|^2+\omega_\phi^2(t)\left|\varphi_p(t)\right|^2 \right]
\end{align}
and 
\begin{align}
\Lambda_\phi(p,t)=\frac{1}{\omega_\phi(t)}\left[\left(\dot{\varphi}_p(t)-\varphi_p (t)\frac{\dot \Omega}{\Omega}\right)^2+\omega_\phi^2(t)\left(\varphi_p(t)\right)^2 \right] .
\end{align}
The functions \(K_\phi\) and \(\Lambda_\phi\) obey the relation: 
\begin{align}
    K_\phi^2(p,t)-\Lambda_\phi(p,t)\Lambda_\phi^*(p,t)=1.
\end{align}
For the conformal theory \(\xi=\frac{1}{6}\) and \(m=0\). Then one can show that the mode function \(\varphi_p(t)=\frac{1}{\sqrt{2 p}}e^{-i |p| t}\) is the positive energy solution for all times. In such a case there should be a unique vacuum state. However, the canonical Hamiltonian operator is not diagonal since \(\Lambda_\phi(p,t) \ne 0\). This tells us that there is something wrong with such a definition of the Hamiltonian. 

One can define the Hamiltonian operator using the stress-energy tensor. The variation of the action with respect to the metric \(g_{\mu\nu}\) gives the corresponding operator:
\begin{gather}
    T_{\mu \nu} = \partial_\mu \phi(x) \partial_\nu \phi(x) - \frac{1}{2} g_{\mu \nu} \left(\partial_\rho \phi(x) \partial^\rho \phi(x) - m^2 \phi^2(x)+\xi R \phi^2(x)\right) \nonumber 
    +\\+\xi R_{\mu \nu }\phi^2(x)+\xi\left(g_{\mu\nu}\Box -\triangledown_\mu \triangledown_\nu\right) \phi^2(x).
\end{gather}
Then the Hamiltonian operator is:  
\begin{gather}
    \hat{E}= \int d^3 x \sqrt{g} \ \hat{T}^0_0
    =\\= \nonumber  
        \frac{1}{2} \int d^3x \sqrt{g} \Bigg( \nonumber  g^{00}\partial_t \hat{\phi}(x)\partial_t \hat{\phi}(x) -g^{i j}\triangledown_i\hat{\phi}(x)\triangledown_j\hat{\phi}(x)+\left(m^2-\xi R\right)\hat{\phi}^2(x)\Bigg)
        + \\+\nonumber
          \xi \int d^3x \sqrt{g} \Bigg( R_0^0 \hat{\phi}^2(x) +\left(\Box -\triangledown^0 \triangledown_0\right) \hat{\phi}^2(x)\Bigg).
\end{gather}
As we can see, the canonical and newly defined Hamiltonians are not equivalent for non-minimally coupled theory. They are equivalent only when \(\xi=0\). 

The expression for the last Hamiltonian via the ladder operators is given by: 
\begin{align}
     \hat{E}=\frac{1}{2}\int \frac{d^3 p}{(2\pi)^3} \omega_E(p,t) \bigg[ K_E(p,t) \hat{a}_p \hat{a}_p^\dagger + \Lambda_E(p,t) \hat{a}_p {\hat{a}_{-p}} \bigg]+h.c.,
\end{align}
where:
\begin{align}
 \omega_E^2(p,t)=p^2+\Omega^2(t) m^2+\left(1-6\xi\right)\left(\frac{\dot{\Omega}(t)}{\Omega(t)}\right)^2,
\end{align}

\begin{align}
K_E(p,t)=\frac{1}{   \omega_E(p,t)}\left[\left|\dot{\varphi}_p(t)\right|^2 +   \omega^2_E(p,t)\left|\varphi_p(t)\right|^2+(6\xi-1)\frac{\dot{\Omega}(t)}{\Omega(t)}\partial_t\left|\varphi_p(t)\right|^2  \right]
\end{align}
and 
\begin{align}
\Lambda_E(p,t)=\frac{1}{   \omega_E(p,t)}\left[\left(\dot{\varphi}_p(t)\right)^2+   \omega^2_E(p,t)\left(\varphi_p(t)\right)^2+(6\xi-1)\frac{\dot{\Omega}(t)}{\Omega(t)}\partial_t \left(\varphi_p(t)\right)^2 \right] .
\end{align}
The functions \(K_E\) and \(\Lambda_E\) obey the relation: 
\begin{align}
    K_E^2(p,t)-\Lambda_E(p,t)\Lambda_E^*(p,t)=1.
\end{align}
For the conformal case \(\xi=\frac{1}{6}\) and \(m=0\), this Hamiltonian operator is diagonal: 
\begin{align}
     \hat{E}=\frac{1}{2}\int \frac{d^3 p}{(2\pi)^3} |p|  \hat{a}_p \hat{a}_p^\dagger+h.c..
\end{align}
But this operator does not lead to the correct equations of motion following from the Heisenberg equations for $\hat{\phi}$ and $\hat{\pi}$. For example, for conformal theory one obtains: 
\begin{align}
    \left[\hat{E},\hat{\phi}\right]=-i \dot{\hat{\phi}}-i \hat{\phi} \frac{\dot{\Omega}(t)}{\Omega(t)}.
\end{align}
The last term on the right hand side appears because the $\phi$ operator depends on time through the presence of $\Omega$. This fact also tells us that there is something wrong with such a definition of the Hamiltonian.


Now let us consider another definition of the Hamiltonian operator. To do that let us rewrite the action in terms of the new canonival variable \(\varphi(x)=\phi(x) \Omega(t)\): 
\begin{gather}
\mathcal{S}=\nonumber\frac{1}{2}\int d^4 x  \Bigg[\partial_t \varphi(x) \partial_t \varphi(x) - \\ - \Omega^2(t)\Bigg(-g^{i j}\partial_i \varphi(x) \partial_j \varphi(x) +m^2\varphi^2(x)+\left(\xi-\frac{1}{6}\right) \frac{6\ddot{\Omega}(t)}{\Omega^3(t)} \varphi^2(x)\Bigg) \Bigg].
\end{gather}
Then the variation of the action with respect to \(\varphi\) gives the same equation of motion as \eqref{mode}. The canonical momentum is now \(\pi(x)= \partial_t \varphi(x)\). Hence, the new canonical Hamiltonian is given by:
\begin{gather}
    \hat{H}_\varphi
    = \frac{1}{2} \int d^3x \Bigg[\hat{\pi}^2(x) + \\ + \Omega^2(t)\Bigg(-g^{i j}\partial_i \hat{\varphi}(x) \partial_j \hat{\varphi}(x) + m^2 \hat{\varphi}^2(x) + \left(\xi-\frac{1}{6}\right) \frac{6\ddot{\Omega}(t)}{\Omega^3(t)} \hat{\varphi}^2(x)\Bigg)\Bigg],
\end{gather}
This definition is more natural from the point of view of the Hamiltonian mechanics since the contribution containing the time derivative of the field operator is not multiplied by the time-dependent function $(\sqrt{g} g^{00})$, as it was the case in the definition \eqref{Ham}. Futhermore this definition of the Hamiltonian leads to the correct equations of motion.

The expression for the canonical Hamiltonian \(H_\varphi\) via the ladder operators is: 
\begin{align}
     \hat{H}_\varphi=\frac{1}{2}\int \frac{d^3 p}{(2\pi)^3} \omega_\phi(p,t) \bigg[ K_\varphi(p,t) \hat{a}_p \hat{a}_p^\dagger + \Lambda_\varphi(p,t) \hat{a}_p {\hat{a}_{-p}} \bigg]+h.c.,
\end{align}
where:
\begin{align}
    \omega_\varphi^2(p,t)=p^2+\Omega^2(t)\left(m^2+\left(\xi-\frac{1}{6}\right)\frac{6\ddot{\Omega}(t)}{\Omega^3(t)}\right),
\end{align}
\begin{align}
K_\varphi(p,t)=\frac{1}{\omega_\varphi(p,t)}\left[\left|\dot{\varphi}_p(t)\right|^2+\omega_\varphi^2(p,t)\left|\varphi_p(t)\right|^2 \right]
\end{align}
and 
\begin{align}
\Lambda_\varphi(p,t)=\frac{1}{\omega_\varphi(p,t)}\left[\left(\dot{\varphi}_p(t)\right)^2+\omega_\varphi^2(p,t)\left(\varphi_p(t)\right)^2 \right] .
\end{align}
The functions \(K_\varphi\) and \(\Lambda_\varphi\) obey the relation: 
\begin{align}
    K_\varphi^2(p,t)-\Lambda_\varphi(p,t)\Lambda_\varphi^*(p,t)=1.
\end{align}
For the conformal case \(\xi=\frac{1}{6}\) and \(m=0\), this canonical Hamiltonian is diagonal since \(\Lambda_\varphi=0\).

All these three Hamiltonian operators \((\hat{H}_\phi,\hat{E},\hat{H}_\varphi)\) for generic values of $\xi$ and $m$ cannot be diagonalized once and for all, because there is no solution to the Klein-Gordon equation that coincides with the function that solves the equation \(\Lambda_{\phi,E,\varphi}(p,t)=0\). However, one can diagonalize them at a particular moment of time \(t=t_i\), using the following Bogolyubov transformation: 
\begin{gather}
    \hat{a}_p=A(p,t_i) \hat{\xi}_{p}(t_i)+B^*(p,t_i) \hat{\xi}^\dagger_{-p}(t_i),
\end{gather}
where \(\hat{\xi}_{p}(t_i)\) and \(\hat{\xi}^\dagger_{-p}(t_i)\) are new ladder operators.
Using the hyperbolic expressions for \(K(p,t_i)\) and \(\Lambda(p,t_i)\):
\begin{align}
    K(p,t_i)=\cosh(2\Theta(p,t_i)) \quad \text{and} \quad \Lambda(p,t_i)=e^{i \xi(p,t_i)} \sinh(2\Theta(p,t_i)), 
\end{align}
one can express the Bogolyubov coefficients \(A(p,t_i)\) and \(B(p,t_i)\) that make \(\Lambda(p,t_i) = 0\) as follows: 
\begin{align}
    A(p,t_i)=\cosh(\Theta(p,t_i)) \quad \text{and} \quad B(p,t_i)=-e^{i \xi(p,t_i)} \sinh(\Theta(p,t_i)).
\end{align}
Thus, the time-dependent Hamiltonian is now diagonalized at \(t=t_i\), but is not diagonal at any other time: 
\begin{align}
    :\hat{H}(t_i):=\int \frac{d^3 p}{(2\pi)^3} \omega(p,t_i) \hat{\xi}^\dagger_{p}(t_i) \hat{\xi}_{p}(t_i).
\end{align}
As a result, in the vicinity of any given moment of time \(t_i\), we can define the Fock space ground state \(|t_i\rangle\), which is annihilated by \(\hat{\xi}_{p}\). This allows us to have a kind of particle interpretation. This state does not contain particles at the particular time \(t_i\). Similarly, one can define a Fock space ground state \(|t_f\rangle\) at another moment of time \(t_f\). The state \(|t_i\rangle\) is no longer the vacuum at \(t_f\). The ladder operators at times \(t_i\) and \(t_f\) are related by the Bogolyubov transformations: 

\begin{gather}
\hat{\xi}_{p}(t_f) = \alpha_p(t_i,t_f)\hat{\xi}_{p}(t_i)+ \beta_p(t_i,t_f)\hat{\xi}^\dagger_{-p}(t_i),
\end{gather}
where:
\begin{gather}
    \alpha_p(t_i,t_f)= A(p,t_i) A(p,t_f)-B(p,t_i)B^*(p,t_f)
\end{gather}
and 
\begin{align}
     \beta_p(t_i,t_f)= B^*(p,t_i) A(p,t_f)-A(p,t_i)B^*(p,t_f).
\end{align}
Hence one can obtain: 
\begin{gather}
|\beta_p(t_i,t_f)|^2= \frac{K(p,t_i)K(p,t_f)-1}{2}-\frac{\Lambda(p,t_i)\Lambda^*(p,t_f)+\Lambda^*(p,t_i)\Lambda(p,t_f)}{4}.
 \end{gather}
Then the probability of finding no particles at time \(t_f\) if one has started from the state \(|t_i\rangle\) is given by:
\begin{gather}
\label{W12}
W_{B}=-\frac{1}{2}\ln | \langle t_f|t_i\rangle|^2
= \\=\nonumber
\frac{V_3}{2} 
 \int \frac{d^3 p}{(2\pi)^3} \ln \left[\frac{K(p,t_i)K(p,t_f)+1}{2}-\frac{\Lambda(p,t_i)\Lambda^*(p,t_f)+\Lambda^*(p,t_i)\Lambda(p,t_f)}{4}\right].
\end{gather}
This quantity is well defined if the integrand decays faster than \(\frac{1}{p^3}\) as \(p \rightarrow \infty\). 

Using the approximate solution in the limit of large momentum\footnote{Such a behavior is assumed from the condition to have the proper Hadamard properties of the propagators, which are built using the corresponding Fock space ground state.}: 
\begin{align}
\label{mode exp}
 \varphi_p(t)\approx   \frac{e^{-i p t}}{\sqrt{2 p}},
\end{align}
one can show that \(|\beta_p(t_i,t_f)|^2\) depends on what definition of the Hamiltonian operator we use to define the vacuum state at a particular moment of time:
\begin{itemize}
    \item canonical Hamiltonian:
\begin{gather}
    |\beta^\phi_p(t_i,t_f)|^2= \\=\nonumber\frac{1}{4p^2}\left(\frac{\dot{\Omega}^2(t_1)}{\Omega^2(t_1)}+\frac{\dot{\Omega}^2(t_2)}{\Omega^2(t_2)}-2 \frac{\dot{\Omega}(t_1)}{\Omega(t_1)} \frac{\dot{\Omega}(t_2)}{\Omega(t_2)} \cos\left(2p(t_1-t_2)\right)\right) +O\left(\frac{1}{p^4}\right).
\end{gather}
\item the Hamiltonian operator following from the stress-energy tensor:
\begin{gather}
    |\beta^E_p(t_i,t_f)|^2= \\=\nonumber \frac{1}{4 p^2}\left(1-6 \xi\right)^2 \left(\frac{\dot{\Omega}^2(t_1)}{\Omega^2(t_1)}+\frac{\dot{\Omega}^2(t_2)}{\Omega^2(t_2)}-2 \frac{\dot{\Omega}(t_1)}{\Omega(t_1)} \frac{\dot{\Omega}(t_2)}{\Omega(t_2)} \cos\left(2p(t_1-t_2)\right)\right)+O\left(\frac{1}{p^4}\right).
\end{gather}
\item new canonical Hamiltonian: 
\begin{align}
    |\beta^\varphi_p(t_i,t_f)|^2 = 0 \times \frac{1}{p^2}+O\left(\frac{1}{p^4}\right).
\end{align}
\end{itemize} 

Therefore, for the first two cases, the integral in \eqref{W12} is divergent. Correspondingly, the number of created particles does diverge for the first two cases: 
\begin{align}
    N=\int \frac{d^3 p}{(2\pi)^3} |\beta_p(t_i,t_f)|^2.
\end{align}
In the last case, \(|\beta_\varphi|^2\) decays faster than \(\frac{1}{p^3}\) as \(p \rightarrow \infty\), but to reproduce the explicit form of the terms \(\frac{1}{p^4}\), we should take into account subleading contributions in the expansion of the mode function \eqref{mode exp}. Hence, in this case, the rate and number of created particles are well defined. This means that the third definition of the canonical Hamiltonian is more natural.

\section{The functional integral approach}

The in-out amplitude is usually written in terms of the functional integral in the following way:
\begin{equation}\label{path cov}
    \langle out|in\rangle= \int{\cal D} \phi \ e^{i\mathcal{S}}=e^{i\mathcal{S}_{eff}}.
\end{equation}
It is straightforward to see that 
\begin{equation}
    \partial_{m^2}\ln \left(\int{\cal D}\phi \ e^{i \mathcal{S}}\right)=\frac{i}{2}\int d^4x \sqrt{|g|}G(x,x),
\end{equation}
where $G$ is the in-out Feynman propagator taken at coincident points:
\begin{align}
    G(x,x) =\frac{\int{\cal D}\phi \phi(x)\phi(x)e^{i\mathcal{S}}}{\int{\cal D}\phi \ e^{i\mathcal{S}}}= \frac{\langle out | \hat{\phi}(x) \hat{\phi}(x) | in \rangle}{\langle out | in \rangle}.
\end{align}
This allows one to express the effective action via the Feynman propagator \cite{Candelas:1975du,Polyakov:2007mm,Akhmedov:2009ta,Akhmedov:2019esv}:
\begin{align}
\label{10}
   \mathcal{S}_{eff} = - \frac{1}{2} \int d^{4} x \sqrt{|g|} \int\limits_{\infty}^{m^2} d \tilde{m}^2 G(x,x).
\end{align}
The propagator is divergent at the coincident points, but its imaginary part is finite and determines the probability of particle creation:
\begin{align}
\label{WP}
    W_P=-\ln | \langle out|in\rangle|=\Im \mathcal{S}_{eff},
\end{align} 
where the index $P$ means that this quantity is obtained from the functional integral \eqref{path cov} or the propagator. In contrast to the method discussed in the previous section, where we use Bogolyubov coefficients to define the imaginary part of the effective action, this method gives the well defined definition of the rate of particle creation, since the imaginary part of the effective action is given by the four-volume integral of the rate as follows: 
\begin{align}\label{4.7}
    w_P(x)= -\frac{1}{2} \int_{\infty}^{m^2} d \tilde{m}^2  \ \Im G(x,x).
\end{align}
Hence, in the case of eternally expanding space-time when the total number of produced particles is divergent, we do not needed to identify the divergent integral over the momentum with the infinite space-time volume, since the volume integral is already present in the definition \eqref{10}. 

The relation \eqref{4.7} is useful only in situations when it is possible to calculate the in-out propagator exactly. An example of such a case is de Sitter space-time. In the next section we will show results for the rate of particle creation in expanding Poincare patch and in global de Sitter space-time and will show that two methods of calculations described above give different answers. At the same time, we will show that the two methods give the same answer in the case when the background field is acting for a finite period of time. 

\section{Poincare patch of de Sitter space-time} 

In the Poincare patch of de Sitter space-time the metric is:

\begin{equation}
\label{ds Poincare}
    d s^2 = d t^2 - e^{2 t} d \vec{x}^2= \eta^{-2}\left(d\eta^2-d\vec{x}^2\right),
\end{equation}
where we set the Hubble constant to one and $\eta=e^{-t}$. This metric is not flat at past and future infinities, but one usually defines the ''in'' and ''out'' modes as follows (see e.g. \cite{Birrell:1982ix}):
\begin{align}
    \varphi^{in}_k= \sqrt{\frac{\pi e^{-\mu\pi}}{4}}\eta^{\frac{1}{2}} H^{(1)}_{i\mu}(k \eta) \quad \text{and} \quad \varphi^{out}_k =\sqrt{\frac{\pi}{2\sinh(\mu\pi)}}\eta^{\frac{1}{2}}J_{i\mu}(k\eta),
\end{align}
where $\mu=\sqrt{m^2-\frac{9}{4}}$.
The ''in'' or Bunch–Davis (BD) modes behave as single oscillating exponents at past infinity: $H \sim e^{i p \eta}$ for $p\eta \gg \mu$. The out–modes behave as single oscillating exponents at future infinity: $J_{i\mu}\sim e^{-i\mu t}$ for $p\eta\ll\mu$. While ''in'' modes do approximately diagonalise the free Hamiltonian at past infinity, $\eta\rightarrow+\infty$, of the Poincare patch, the ''out'' modes do not diagonalise the Hamiltonian at any time. In fact, the in-out amplitude here has a quite different physical meaning than in situations with well-defined ``in'' and ``out'' states. However, it is still possible to write such an amplitude via the functional integral and check if two approaches described above give the same answer or not. 

The ''in'' and ''out'' modes are related by the Bogolyubov transformation \eqref{bogolyubov} with $ \alpha_{p} = \left(1-e^{-2\pi \mu}\right)^{-\frac{1}{2}}$ and  $\beta_{p} = -\left( e^{2 \pi \mu} - 1\right)^{-\frac{1}{2}}$. Hence, the probability of particle creation calculated in the first approach is given by:
\begin{equation}
\label{W_B poincare}
    W_B = - \frac{V_{3}}{4} \int\frac{d^3 p}{(2\pi)^3} \ln{\left(1-e^{-2\pi\mu}\right)}.
\end{equation} 
The integral over $p$ in \eqref{W_B poincare} is divergent. 
 Let's cut the integration over $p$ by $\Lambda$, to obtain that $\int \frac{d^3p}{(2\pi)^3}= \frac{\Lambda^3}{6\pi^2} $. Then for a fixed physical momentum cutoff $p_{phys}=\Lambda e^{-t}$, an increase in time by $\delta t$ results in the such an increase in $\Lambda$ that: $\delta \Lambda =\Lambda \delta t$ as $\Lambda$ and $t$ are taken to infinity.  As a result one can associate the divergent integral over time with the divergent integral over $p$: 
\begin{align}
     \delta V_4=\int d^4 x \sqrt{g}\Big|_{t}^{t+\delta t}=V_3 \Lambda ^2 \delta \Lambda =  2\pi^2 V_3 \delta \int \frac{d^3p}{(2\pi)^3}.
\end{align} 
As a result the probability of the particle creation is given by: 
\begin{equation}
    W_B = - \frac{V_{4}}{4}H^4 \ln{\left(1-e^{-2\pi\mu}\right)},
\end{equation}
where we restore the Hubble constant.

Another method of computation of the probability of particle creation uses the imaginary part of Feynman ''in-out'' propagator. This propagator is de Sitter invariant \cite{Polyakov:2007mm,Akhmedov:2019esv}. As a result, it's imaginary part at coincident points does not depend on the space-time position:   

\begin{equation}
\label{propagatorpoincare}
    \Im G(x,x) = \frac{(1/4 + \mu^2)e^{-\pi\mu}}{16\pi\cosh{(\mu\pi)}}.
\end{equation}
Therefore the probability of the particle creation is given by: 
\begin{gather}
    W_P = - \frac{1}{2} \int d^{4} x \sqrt{|g|} \int_{\infty}^{m^2} d \tilde{m}^2 \Im G(x,x)
    =\\=
    \nonumber
    \frac{1}{2} V_4 \int_{\infty}^{m^2} d m^2  \frac{(1/4 + \mu^2)e^{-\pi\mu}}{16\pi\cosh{(\mu\pi)}},
\end{gather}
where the space-time volume naturally factors out, in contrast to the previous method. The last integral can be evaluated analytically in terms of polylogarithms, but to simplify the discussion let us consider here only the large mass limit. In this approximation the probability of the particle creation is given by:
\begin{equation}
    W_P \approx V_4\frac{H^4}{16 \pi^{2}} \left(\frac{m}{H}\right)^{3} e^{- 2 \pi m/H}, \hspace{0.25cm} m \gg H.
\end{equation}
As one can see, the two methods give different answers.  And only in the second method, the dependence on the space-time volume naturally factors out from the definition rather than from a rather unnatural сut-off procedure. Such a difference appears also in calculations in other patches of the de Sitter space-time, as will be shown below, and for example in the future Rindler wedge \cite{Akhmedov:2021agm}. However on the example of the Poincare patch we will explain the reason of such a difference at the end of our paper.

\section{ Global de Sitter space-time } 

Global de Sitter space-time metric is as follows:
\begin{equation}
\label{ds Global}
    d s^2 = d t^2 - \cosh^{2} t \  d \Omega ^2,
\end{equation}
where we set the Hubble constant to one. 

This metric is not flat at past and future infinities. But one usually defines ''in'' and ''out'' modes as follows \cite{Mottola:1984ar}:
\begin{align}
    \varphi^{in}_{j m}= \sqrt{\frac{\pi}{2\sinh \pi \mu}}\cosh^{-3/2}(t) \mathrm{Q}^{-i
    \mu}_{j+1/2}(\tanh t) Y_{j m}(\Omega)
\end{align}
and
\begin{align}
    \varphi^{out}_{j m}= \sqrt{\frac{\pi}{2\sinh \pi \mu}}\cosh^{-3/2}(t) \mathrm{P}^{-i
    \mu}_{j+1/2}(\tanh t) Y_{j m}(\Omega).
\end{align}
The ''in''-modes behave as single oscillating exponents at past infinity: $\varphi^{in}_{j m} \sim e^{-i \mu t}$, $\eta = e^{-t}$. The ''out''–modes behave as single oscillating exponents at future infinity: $ \varphi^{out}_{j m} \sim e^{-i\mu t}$. While ''in'' and ''out'' modes behave as single exponents at past and future infinities, they do not diagonalise the free Hamiltonian at any time. As a result we confront with the same problem as is Poincare patch.

The ''in'' and ''out'' modes are related by the Bogolyubov transformations \eqref{bogolyubov} with $\alpha_{j} = - i \coth{(\mu \pi)}   $ and  $\beta_{j} = \frac{\Gamma(j+3/2-i\mu)}{\Gamma(j+3/2+i\mu)\sin{(i\mu\pi)}} $. Hence the probability of the particle creation in the first method is given by:
\begin{equation}
    W_B =  \frac{1}{2} \sum\limits_{j,m} \ln{\coth{(\mu\pi)}}.
\end{equation} 
As in the previous section, we are faced with a divergent expression. Let's cut the summation over $j$ by a large number $N$, so that $\sum_{j,m}1\sim \frac{N^3}{3}$ for $N\gg1$. Then for a fixed physical momentum cutoff $p_{phys}=\frac{N}{\cosh t}$, an increase in time by $\delta t$ results in such an increase in $N$ that: $\delta N =N \delta t$, as $N$ and $t$ are taken to infinity.  As a result one can associate the divergent integral over time with the divergent sum over $j$: 
\begin{align}
     \delta V_4=\int d^4 x \sqrt{g}\Big|_{t}^{t+\delta t}=\frac{\pi^2}{4}N^2 \delta N = \frac{\pi^2}{4} \delta \sum_{j,m}1. 
\end{align} 
Hence the probabilty of the particle creation is given by \cite{Mottola:1984ar}:
\begin{equation}
    W_B = V_4\frac{2}{\pi^2} \ln{\coth{(\mu \pi)}}.
\end{equation}

Now let us look at the second method. Again, because the in-out propagator is de Sitter invariant, the imaginary part at coincident points does not depend on the space-time position. This time it is equal to:   

\begin{equation}
\label{propagatorpoincare1}
    \Im G(x,x) =  - \frac{(1/4+\mu^2)}{16 \pi \cosh^{2}{(\mu \pi)}}.
\end{equation}
Therefore the probability of the particle creation is given by: 
\begin{gather}
    W_P = - \frac{1}{2} \int d^{4} x \sqrt{|g|} \int_{\infty}^{m^2} d \tilde{m}^2 \Im G(x,x)
    =\\=
    \nonumber
    \frac{1}{2} V_4 \int_{\infty}^{m^2} d m^2 \frac{(1/4+\mu^2)}{16 \pi \cosh^{2}{(\mu \pi)}},
\end{gather}
The last integral can be evaluated analytically in terms of polylogarithms. In the large mass limit the probability of the particle creation is given by:
\begin{equation}
    W_P \approx  V_4\frac{H^4}{8 \pi^{2}} \left(\frac{m}{H}\right)^{3} e^{-2 \pi m/H}.
\end{equation}
Again the two methods give different answers. As we will show below, this is related to the fact that one has to correct the relation (\ref{path cov}).  To do that let us look closer at the procedure of the derivation of the functional integral and show that in situations where background fields act for finite periods of time the two methods give the same result. 

\section{On the functional integral in external fields.}

                We want to express the in-out amplitude as the functional integral. We briefly repeat the procedure which is written in details in \cite{Weinberg:1995mt} and extend it to the case of the presence of an external field.
                
            One can write down the in-out amplitude as the following functional integral:
            \begin{gather}
                    \bra{out}\ket{in}=\int{\cal D}\phi(x,t)e^{i\int^{+\infty}_{-\infty}dt L[\phi(x,t),\dot{\phi}(x,t),t]}\Psi_{out}\Psi^{*}_{in},
            \end{gather}
            where vacuum functionals are defined as follows: 
            \begin{align}
                    \Psi_{out}=\langle out | \phi(x,+\infty)\rangle\quad \text{and} \quad \Psi^{*}_{in}=\langle\phi(x,-\infty)| in\rangle.
            \end{align}
              Obviously the expressions in the last equations are defined as limiting values at $t\rightarrow\pm\infty$. We have mentioned that the calculation of the particle creation probability through imaginary part of the in-out propagator reproduces the answer obtained via Bogolyubov coefficients only in the case then the background field is acting for a finite time. In principle in such a calculation one should take into account the fact that vacuum functionals, $\Psi_{in}$ and $\Psi_{out}$, are also mass-dependent.
              
              Using the definitions of the in and out states one can write the functional equation for the vacuum functionals $\Psi_{in}$ and $\Psi_{out}$. For example, consider the in state:
              \begin{equation}\label{INdef}
                  \begin{aligned}
                      & \hat{a}^{in}_p|in\rangle=0,
                  \end{aligned}
              \end{equation}
        where $\hat{a}^{in}_p$ is defined in eq. \eqref{fieldoperator} and we assume that $\underset{t\rightarrow-\infty}{\lim}\omega_p(t)=\omega_{p-}=const$, i.e. the background field is off at the past infinity. Then eq.  \eqref{INdef} can be rewritten as:
            \begin{equation}
                \int d^3x \, e^{ipx} \, \left[\frac{\dot{\varphi}^{in*}_{p}(t)}{\varphi^{in*}_{p}(t)}\phi(x,t)+i\frac{\delta}{\delta\phi(x,t)}\right]\Psi_{in}=0.
            \end{equation}
            It is straightforward to show that:
            \begin{gather}
                    \Psi_{in}\propto \exp\left[i\frac{1}{2}\int d^3xd^3y\phi(x,t)\phi(y,t)E(x,y)\right],
            \end{gather}
            where
            \begin{align}
            \label{E(x,y)}
                E(x,y)=\int \frac{d^3p}{(2\pi)^3}e^{ip(x-y)}\frac{\dot{\varphi}^{in*}_{p}(t)}{\varphi^{in*}_{p}(t)}.
            \end{align}
Now, using \eqref{inassym}, one can take the limit \( t \rightarrow -\infty \), to obtain:
\begin{gather}
\Psi_{in} = \prod_p \left( \frac{\omega_{p-}}{\pi} \right)^{\frac{1}{4}} \exp \left[ -\frac{1}{2} \int d^3x \, d^3y \, \phi(x, -\infty) \phi(y, -\infty) \mathcal{E}(x, y) \right],
\end{gather}
where we explicitly find the normalization condition and define the limit of the function \eqref{E(x,y)} as follows:
\begin{align}
\mathcal{E}(x, y) = \int \frac{d^3p}{(2\pi)^3} e^{ip(x-y)} \omega_{p-}.
\end{align}
Logarithmic derivative of vacuum functional $\Psi_{in}$ over $m^2$ gives:
            \begin{gather}
                \label{psiinderiv}
                   \frac{\int{\cal D}\phi(x,t)e^{i\int^{+\infty}_{-\infty}dt L[\phi(x,t),\dot{\phi}(x,t),t]}\Psi_{out}\partial_{m^2}\Psi^{*}_{in}}{\bra{out}\ket{in}}
                   =\\
                   \nonumber
                    =V_3\int\frac{d^3p}{(2\pi)^3} \frac{1}{8\omega^2_{p-}}-\int\frac{d^3p}{(2\pi)^3}\int d^3xd^3y\frac{1}{4\omega_{p-}}e^{ip(x-y)} \langle \phi(x,-\infty)\phi(y,-\infty)\rangle,
    \end{gather}
    where
   \begin{gather}
                           \langle\phi(x,-\infty)\phi(y,-\infty)\rangle=G_{in-out}(x,-\infty,y,-\infty).
            \end{gather}
            Using the mode expansion of the field operator, the second integral on the RHS of (\ref{psiinderiv}) can be rewritten as:
            \begin{gather}\label{7.7}
                                    \int\frac{d^3p}{(2\pi)^3}\int d^3xd^3y\frac{1}{4\omega_{p-}}e^{ip(x-y)}\langle \phi(x,-\infty)\phi(y,-\infty)\rangle
                                    =\\= \nonumber
                                    V_3\int\frac{d^3p}{(2\pi)^3}\frac{1}{4\omega_{p-}}\frac{\varphi^{in*}_{p}(-\infty)\varphi^{out}_{p}(-\infty)}{\alpha^{*}_p}
                                    =\\= \nonumber
                                    V_3\int\frac{d^3p}{(2\pi)^3}\frac{1}{4\omega_{p-}}\left[\varphi^{in*}_{p}(-\infty)\varphi^{in}_{p}(-\infty)-\frac{\beta^{*}_p}{\alpha^{*}_p}\varphi^{in*}_{p}(-\infty)\varphi^{in*}_{p}(-\infty)\right]
                                    =\\=
                                    \nonumber
                                    V_3\int\frac{d^3p}{(2\pi)^3}\frac{1}{8\omega^2_{p-}}-\underset{t\rightarrow-\infty}{\lim}V_3\int\frac{d^3p}{(2\pi)^3}\frac{1}{8\omega^2_{p-}}\frac{\beta^{*}_p}{\alpha^{*}_p}e^{2i\omega_{p-}t}.
            \end{gather}
            The first term in the last line of (\ref{7.7}) cancels the first term on the RHS of (\ref{psiinderiv}). A similar term will arise from the out-vacuum functional. Combining all these observations together, one obtains that:
            \begin{gather}
                    \partial_{m^2} \ln \left(\int{\cal D}\phi(x,t)e^{i\int^{+\infty}_{-\infty}dt L[\phi(x,t),\dot{\phi}(x,t),t]}\Psi_{out}\Psi^{*}_{in}\right)
                    =\\ = \nonumber
                    \underset{t\rightarrow-\infty}{\lim}V_3\int\frac{d^3p}{(2\pi)^3}\frac{1}{8\omega^2_{p-}}\frac{\beta^{*}_p}{\alpha^{*}_p}e^{2i\omega_{p-}t}-\underset{t\rightarrow+\infty}{\lim}V_3\int\frac{d^3p}{(2\pi)^3}\frac{1}{8\omega^2_{p-}}\frac{\beta_p}{\alpha^{*}_p}e^{-2i\omega_{p+}t}
                    -\\ -\nonumber
                    i\frac{1}{2}\int d^4xG_{in-out}(x,t,x,t).
            \end{gather}
        Again we can use the mode expansion for field operator to obtain that:
        \begin{equation}\label{inttt}
            \int d^4x\ G_{in-out}(x,t,x,t)=V_3\int dt \int\frac{d^3p}{(2\pi)^3}\frac{1}{\alpha^{*}_p}\varphi^{in*}_{p}(t)\varphi^{out}_{p}(t)
        \end{equation}
To make further tranformations we use the following relation:
\begin{gather}
        \varphi^{in*}_{p}(t,m)\varphi^{out}_{p}(t,m)=\lim_{\bar{m}\to m}\frac{\bar{m}^2-m^2 }{ \bar{m}^2-m^2  } \varphi^{in}_{p}(t,\bar{m})\varphi^{out}_{p}(t,m)
        =\\ = \nonumber
        - \partial_t \Big[ \partial_t \partial_{m^2} \varphi^{in}_{p}(t,m)\varphi^{out}_{p}(t,m)-  \partial_{m^2} \varphi^{in}_{p}(t,m)\partial_t \varphi^{out}_{p}(t,m)\Big],
    \end{gather}
where $ \varphi^{in}_{p}(t,\bar{m})$ and $ \varphi^{out}_{p}(t,m)$ solve the equations of motions \eqref{mode} with the corresponding masses.

Because integrals appearing on the RHS of (\ref{inttt}) are convergent (for the background field that acts for a finite period of time), we can change the order of integration and perform first the integration over time. After some simplifications one obtains that:
            \begin{gather}
            \int d^4xG_{in-out}(x,t,x,t)
            =\\= \nonumber
            -iV_3\int\frac{d^3p}{(2\pi)^3}\left[\frac{\partial_m^2\alpha^{*}_p}{\alpha^{*}_p}\right]+i\underset{t\rightarrow+\infty}{\lim}V_3\int\frac{d^3p}{(2\pi)^3}\left[\frac{1}{4\omega_{p+}}\frac{\beta}{\alpha^{*}_p}e^{-2i\omega_{p+}t}
            - \frac{it}{4\omega_{p+}}\right]
                    -\\- \nonumber
                    i\underset{t\rightarrow-\infty}{\lim}V_3\int\frac{d^3p}{(2\pi)^3}\left[\frac{1}{4\omega^2_{p-}}\frac{\beta^{*}_p}{\alpha^{*}_p}e^{2i\omega_{p-}t}+\frac{it}{4\omega_{p-}}\right].
                \end{gather}
Oscillating terms cancel each other and finally we obtain that:

            \begin{gather}
            \label{finitfinal}
                    \partial_{m^2} \ln \left(\int{\cal D}\phi(x,t)e^{i\int^{+\infty}_{-\infty}dt L[\phi(x,t),\dot{\phi}(x,t),t]}\Psi_{out}\Psi^{*}_{in}\right)
                    =\\=\nonumber
                    -\frac{V_3}{2}\int\frac{d^3p}{(2\pi)^3}\left[\frac{\partial_m^2\alpha^{*}_p}{\alpha^{*}_p}\right]-i\underset{t\rightarrow+\infty}{\lim}V_3\int\frac{d^3p}{(2\pi)^3}\frac{t}{8\omega_{p+}}+i\underset{t\rightarrow-\infty}{\lim}V_3\int\frac{d^3p}{(2\pi)^3}\frac{t}{8\omega_{p-}}.
                \end{gather}
            The last two terms do not contribute to the imaginary part of the effective action, while the first term reproduces the answer calculated via the Bogolyubov coefficients.

 Thus, when the background field acts for a finite period of time imaginary part of the effective action depends only on the Bogolyubov coefficients:
\begin{align}
\label{ImSeff}
  W_P =\Im \mathcal{S}_{eff}=\frac{V_3}{2} 
 \int \frac{d^3 p}{(2\pi)^3} \ln |a_p|,
\end{align}
and we obtain that  $W_P=W_B$.

In cases when background fields act for infinite periods of time the probability of particle creation divided by $V_3$ is expected to be infinite. Hence one cannot safely change the order of integration\footnote{In the context of space-times with Killing horizons, a similar issue arises when computing the functional integral \cite{Diakonov:2023jdk}.} over the momentum and over time in the integral \eqref{inttt}. This gives a hint why the two methods of the calculation of the effective action can give different answers. In the next section we will look closer to the situation on the example of the Poincare expanding patch of de Sitter space time and explain the reason why the two methods of calculation give different answers.

\section{The functional integral in the expanding Poincare patch}

In the expanding Poincare patch the background gravitational field acts for infinite time. In this case the procedure to construct the functional integral is the same, but there is a difference, which arises due to vacuum functionals, $\Psi_{in}$ and $\Psi_{out}$. The in-vacuum is well defined, see e.g. \cite{Akhmedov:2024ice}. Essentailly because of the latter fact it is straightforward to show that the in-vacuum functional does not contribute to the imaginary part of the effective action\footnote{To make the story shorter we do not give details of this calculation. Essentially for the BD state it is possible to take the limit $t\to -\infty$ without problems that arrise for the ``out'' vacuum functional. As a result the ``in'' vacuum functional is mass independent.}. 

At the same time there are problems that appear due to the out-vacuum functional, which are related to the fact that the background field is not switched off at future infinity. In all, we are interested only in the contribution of the vacuum functional for out-state which is defined as:
                \begin{gather}
                    d_p|out\rangle=0,\quad
            \phi(x,\eta)=\int\frac{d^3p}{(2\pi)^3}[d_p\varphi^{out}_{p}(\eta)e^{ipx}+h.c],\\
                \varphi^{out}_{p}(\eta)=\sqrt{\frac{\pi}{2sh(\mu\pi)}}\eta^{\frac{3}{2}}J_{i\mu}(p\eta).
                \end{gather}
Again from these relations one obtains that:
            \begin{equation}
                \int d^3x \, e^{ipx} \, \left[\frac{\dot{\varphi}^{out*}_{p}(\eta)}{\varphi^{out*}_{p}(\eta)}\phi(x,\eta)+i\frac{\delta}{\delta\phi(x,\eta)}\right]\Psi=0,
            \end{equation}
       and then
            \begin{gather}
                \label{limit}
                    \Psi_{out}\propto \underset{\eta\rightarrow0}{\lim} \exp\left[i\int d^3x \, d^3y \, \phi(x,\eta)\phi(y,\eta)E(x,y)\right],\quad \text{where}\\
                    E(x,y)=\int \frac{d^3p}{(2\pi)^3}e^{ip(x-y)}\Omega(\eta)\frac{\dot{\varphi}^{out*}_{p}(\eta)}{\varphi^{out*}_{p}(\eta)}=\int \frac{d^3p}{(2\pi)^3}e^{ip(x-y)}\left(\frac{3}{2\eta^3}+\frac{\dot{J}^{*}_{i\mu}(p\eta)}{\eta^2J^{*}_{i\mu}(p\eta)}\right).
                \end{gather}
In contrast with the situation with the background field acting for a finite time in this case, the future infinity limit, $\eta\rightarrow 0$, does not exist, due to mixing of momentum, $p$, and time, $\eta$, in the mode functions. Now if one calculates the imaginary part of the effective action via the derivative over $m^2$, then one has to take into account terms which arise from the vacuum functional $\Psi_{out}$. As a result:
     \begin{equation}
                \partial_{m^2} \ln\left(\int{\cal D}\phi(x,\eta)e^{i\int^{+\infty}_{-\infty}d\eta L[\phi(x,\eta),\dot{\phi}(x,\eta),\eta]}\Psi_{out}\Psi^{*}_{in}\right)\neq -i\frac{1}{2}\int d^4xG_{in-out}(x,\eta,x,\eta).
  \end{equation}
For example, the following terms arise from the out-vacuum functional:
        \begin{gather}
            \frac{\int{\cal D}\phi(x,t)e^{i\int^{+\infty}_{-\infty}dt L[\phi(x,t),\dot{\phi}(x,t),t]}\partial_{m^2}\Psi_{out}\Psi^{*}_{in}}{\bra{out}\ket{in}}
            =\\= \nonumber 
            V_3\underset{\eta\rightarrow0}{\lim}\int \frac{d^3p}{(2\pi)^3}\partial_m^2\left(\frac{\dot{J}^{*}_{i\mu}(p\eta)}{\eta^2J^{*}_{i\mu}(p\eta)}\right)\frac{(H^{(1)}_{i\mu}(p\eta))^{*}J^{*}_{i\mu}(p\eta)}{\alpha^{*}_p}
            +\\ + \nonumber
            \text{terms from normalization constant of the out-vacuum functional}
            \end{gather}
        Presence of such a derivative over $m^2$ following from the vaccum functions makes impossible to define the in-out functional integral in situations when background fields are not switched off. Furthermore, the absence of terms appearing from the vacuum functionals is the main reason why the two methods of calculation that we consider here give the same answer.

\section{Conclusion}
We have explored two ways to calculate the imaginary part of  effective actions in background gravitational fields. While, when the background field is acting for a finite time, both methods lead to the same result, in cases of eternal external fields this is no longer the case. Example of the latter situation are presented by various patches of the de Sitter space. We show that the discrepancy in the answers in the de Sitter space-time is due to the fact that when calculating the effective action via the functional integral the contributions of the vacuum functionals are not taken into account.

In our oppinion these considerations prove once again that it is worth working very carefully with the functional integral and in-out Feynman technique in background fields. As one can see there can be many non-obvious assumptions. In the end, the correct approach is to consider quantum fields (for generic states and background fields) with the use of the Keldysh-Schwinger technique and to calculate such observables as e.g. energy-momentum tensor, rather than the number of particles or cross-sections (in the situations when there are no asymptotic states).

\section*{Acknowledgments}
Sections 1-3 and 7-8 were performed by Akhmedov E.T., Belkovich I.A. and Kazarnovskii K.A.. Their work was supported by Russian Science Foundation (Project Number: 23-22-00145). Sections 4-6 were performed by Diakonov D.V. and was partially funded within the state assignment of the Institute for Information Transmission Problems of RAS and  by the grants of the Foundation for the Advancement of Theoretical Physics and Mathematics “BASIS”.

\bibliographystyle{unsrturl}
\bibliography{bibliography.bib}
\end{document}